\documentclass[aps,11pt,nofootinbib,preprintnumbers,showpacs]{revtex4}

\usepackage{amssymb, amsfonts, amsmath, bm}
\usepackage{amsfonts}
\usepackage{ulem}
\usepackage[dvipdfmx]{graphicx,color}

\begin{document}




\title{
Multi-charged black lens 
}
\vspace{2cm}

\author{Shinya Tomizawa${}^{1}$\footnote{tomizawasny`at'stf.teu.ac.jp}  
} 
\vspace{2cm}
\affiliation{
${}^1$ Department of Liberal Arts, Tokyo University of Technology, 5-23-22, Nishikamata, Otaku, Tokyo, 144-8535, Japan
}




\begin{abstract} 
We construct an asymptotically flat, stationary and biaxisymmetric supersymmetric black lens solution in five-dimensional $U(1)^3$ supergravity.
It is shown that the spatial cross section of the horizon is topologically the lens space of $L(n,1)$, and the spacetime is regular on/outside the event horizon. 
The black lens carries $(3n+2)$ physical quantities, three electric charges, two angular momenta and $3(n-1)$ magnetic fluxes.
\end{abstract}
\date{\today}
\maketitle




\section{Introduction}\label{sec:1}
It was shown in Refs.~\cite{Cai:2001su,Galloway:2005mf,Hollands:2007aj,Hollands:2010qy} that for an asymptotically flat, stationary and bi-axisymmetric five-dimensional black hole spacetime, the spatial cross section of each connected component of the event horizon must be topologically either a sphere $S^3$, a ring $S^1\times S^2$, or lens spaces $L(p,q)$ $(p,q:$ coprime integers) under the dominant energy condition. 
 Concerned with the first two cases, one has already known the exact solutions in five-dimensional Einstein theory~\cite{Myers:1986un,
Emparan:2001wn,Pomeransky:2006bd}. 
 Nevertheless, as for the third case, 
a few authors~\cite{Evslin:2008gx,Chen:2008fa,Tomizawa:2019acu} attempted to construct an exact solution to the five-dimensional vacuum Einstein equation by using the inverse scattering method, 
but it turned out that the resultant solution always suffers from naked singularities.

\medskip
The supersymmetric solutions provide us various useful information on the corresponding vacuum solutions. 
In particular, as for asymptotically flat supersymmetric black objects in the five-dimensional minimal supergravity (the constructions of these  solutions are based on the scheme developed by Gauntlett {\it et. al}~\cite{Gauntlett:2002nw}), the properties have been so far studied by many authors.
For instance, the possible topologies of the supersymmetric black holes must be either a sphere $S^3$, a ring $S^1\times S^2$, a torus $T^3$ or quotient thereof~\cite{Reall:2002bh}. 
 For the spherical topology,  Breckenridge {\it et al.} first found a supersymmetric black hole with equal angular momenta~\cite{Breckenridge:1996is}. The supersymmetric  black ring with the topology $S^1\times S^2$ has been found in \cite{Elvang:2004rt}.
Furthermore, 
recently, the asymptotically flat supersymmetric black hole solution with the special horizon topology of $L(2,1)=S^3/{\mathbb Z}_2$ has been constructed  by Kunduri and Lucietti~\cite{Kunduri:2014kja} (see \cite{Kunduri:2016xbo}  for the extension to ${\rm U}(1)^3$ supergravity). 
 Subsequently, the black lens solution with more general lens space topologies $L(n,1)=S^3/{\mathbb Z}_n\ (n\ge 3)$ was constructed  in~\cite{Tomizawa:2016kjh} and it was also immediately generalized to the solution with multiple horizons in~\cite{Tomizawa:2017suc}.

\medskip
In this paper, we generalize the asymptotically flat supersymmetric black lens solution~\cite{Tomizawa:2016kjh}  with the horizon topology of $L(n,1)$ in the five-dimensional minimal supergravity to the supersymmetric solution in the five-dimensional $U(1)^3$ supergravity, 
whereas this is also the extension of the black lens solution~\cite{Kunduri:2016xbo} with the horizon topology $L(2,1)$ to the more general lens spaces $L(n,1)\ (n\ge 3)$.
Our strategy to construct such a solution is to consider the Gibbons-Hawking space as a hyper-K\"ahler base space and allow the harmonic functions to have $n$ point sources with appropriate coefficients. 
By imposing suitable boundary conditions on the solution, we consider the configuration of the point sources in which the cross section of the horizon becomes $L(n,1)$ and the spatial infinity  becomes $S^3$. 
The black lens spacetime possesses a nontrivial domain of outer communication
by the existence of $(n-1)$ nontrivial 2-cycles supported by magnetic fluxes outside the horizon. 
One of these 2-cycles has disk topology $D^2$, whereas the others have the topology of $S^2$.

\medskip
We organize the remaining sections of this paper as follows. 
In Sec.~\ref{sec:solution}, we present the the supersymmetric solution of black lenses admitting stationarity and bi-axisymmetry in the five-dimensional $U(1)^3$ supergravity. 
In  Sec.~\ref{sec:boundary}, we impose the boundary conditions on the solution in order that the spacetime is asymptotically flat, has no curvature/conical singularities in the domain of outer communication and no orbifold/ Dirac-Misner string singularities on the axis.   
In Sec.~\ref{sec:analysis}, we compute the conserved charges and discuss the physical aspects of such a black lens. 
In Sec.~\ref{sec:analysis}, we devote ourselves to the summary and discussion on our results.




\section{Black lens solution}
\label{sec:solution}
In the bosonic sector of the five-dimensional $U(1)^3$ supergravity,   the supersymmetric solution can be written as a timelike fiber bundle over a
hyper K\"ahler space~\cite{Gauntlett:2004qy}.  In this work, we choose such a space as the Gibbons-Hawking space, whose metric can be written in spherical polar coordinates $(r,\theta,\phi)$ on ${\mathbb E}^3$ as
\begin{eqnarray}
ds_M^2&=&H^{-1}(d\psi+\chi)^2+Hds^2_3,\quad d\chi=*dH,\label{eq:GH}\\
ds^2_{3}&=&dr^2+r^2d\theta^2+r^2\sin^2\theta d\phi^2, \label{eq:ds3}
\end{eqnarray}
where $H$ is a harmonic function on ${\mathbb E}^3$ and $*$ denotes the Hodge duarity on  ${\mathbb E}^3$.  
When we assume that $\partial/\partial\psi$ is also a Killing vector for the five-dimensional metric,  
the metric, gauge potentials $A^I$ of Maxwell fields and scalar fields $X^I (I=1,2,3)$ with the constraint $X^1X^2X^3=1$ for the supersymmetric solutions are expressed as
\begin{eqnarray}
ds^2&=&-f^2(dt+\omega)^2+f^{-1}ds^2_M,\\
A^I&=&\frac{1}{3}H_I^{-1}(dt+\omega)+\frac{1}{2}\left[K^IH^{-1}(d\psi+\chi)+\xi^I\right],\\
X^I&=&H_I^{-1}(H_1H_2H_3)^{\frac{1}{3}},
\end{eqnarray}
where the functions $f^{-1}$, $H_I\ (I=1,2,3)$ and $1$-forms $\omega,\xi^I$ are given by
\begin{eqnarray}
f^{-1}&=&3(H_1H_2H_3)^{\frac{1}{3}},\\
H_I&=&L_I+\frac{1}{24}H^{-1}|\epsilon_{IJK}|K^JK^K,\\
\omega&=&\omega_\psi(d\psi+\chi)+\hat\omega,\\
d\xi^I&=&-*dK^I\label{eq:dxi}
\end{eqnarray}
with
\begin{eqnarray}
\omega_\psi&=&-\frac{K^1K^2K^3}{8H^2}-\frac{3L_IK^I}{4H}+M,\\
*d\hat\omega&=&HdM-MdH-\frac{3}{4}(K^IdL_I-L_IdK^I). \label{eq:domega}
\end{eqnarray}
$K^I$, $L_I$ and $M$ are the harmonic functions on ${\mathbb E}^3$. 
We assume that the harmonic functions $H$, $K^I$, $L_I$ $(I=1,2,3)$ and $M$  have $n$ point sources at ${\bm r}=(0,0,z_i)$\ ($z_i<z_j$ for $i<j$, $i,j=1,\ldots,n$) on ${\mathbb E}^3$ as
\begin{eqnarray}
H&=&\frac{h_i}{r_i}:=\frac{n}{r_1}-\sum_{i=2}^n\frac{1}{r_i},
\end{eqnarray}
\begin{eqnarray}
K^I=\sum_i\frac{k^I{}_i}{r_i}, \quad L_I=l_{I0}+\sum_i\frac{l_{Ii}}{r_i},\quad M=m_{0}+\sum_i\frac{m_{i}}{r_i},
\end{eqnarray}
where $r_i=\sqrt{r^2-2rz_i\cos\theta+z_i^2}$.
It can be shown from Eqs.(\ref{eq:GH}), (\ref{eq:domega}) and (\ref{eq:dxi}) that the $1$-forms ($\chi,\hat \omega,\xi^I$) are explicitly   expressed as
\begin{eqnarray}
\chi&=&\sum_ih_i \frac{r\cos\theta -z_i}{r_i}d\phi,\\
\hat\omega&=&\sum_{i,j}\left(h_im_j-\frac{3}{4}\sum_Ik^I{}_il_{Ij}\right)\frac{r^2-(z_i+z_j)r\cos\theta +z_iz_j}{z_{ji}r_ir_j}d\phi\notag \\
&&-\sum_i\left(m_0h_i-\frac{3}{4}\sum_Ik^I{}_il_{I0}\right)\frac{r\cos\theta -z_i}{r_i}d\phi+cd\phi,\\
\xi^I&=&-\sum_ik^I{}_i \frac{r\cos\theta -z_i}{r_i}d\phi,
\end{eqnarray}
where $c$ is a constant  and $z_{ji}:=z_j-z_i$.




\section{Boundary conditions}\label{sec:boundary}




We require that (i) the spacetime is asymptotically flat at infinity $r\to \infty$, 
(ii) the point, ${\bm r}={\bm r}_1$, at which the harmonic functions $H,K^I,L_I,M$ diverge, corresponds to a smooth degenerate Killing horizon whose spatial topology is the lens space of $L(n,1)$, 
and
(iii) the remaining $(n-1)$ points ${\bm r}={\bm r}_i\ (i=2,\ldots,n)$ of the harmonic functions behave as regular points.
We also demand the regularity conditions that there exist no curvature singularities, no conical singularities and no Dirac-Misner string singularities. 
In addition, we impose the absence of closed timelike curves (CTCs) in the domain of outer communication.
In order that the $(8n+5)$ parameters $(k^I{}_i, l_{I0}, l_{Ii},m_0,m_i,z_i,c)$ satisfy all of these boundary conditions, 
we will derive the constraint equations in what follows.

\subsection{Infinity}
\label{sec:bc_infinity}
For $r\to\infty$, the function $f^{-1}$ and $1$-form $\omega$ behave, respectively, as
\begin{eqnarray}
f^{-1}&\simeq&3(l_{10}l_{20}l_{30})^{\frac{1}{3}},\\
\omega &\simeq & \left(m_0-\frac{3}{4}\sum_{i,I}k^I{}_i l_{I0}\right)\left(d\psi +\cos\theta d\phi \right)-\sum_{i}\left(m_0h_i-\frac{3}{4}\sum_Ik^I{}_i l_{I0}\right)\cos\theta d\phi \notag \\
                    &&+\left(\sum_{i,j(i\not =j)}\frac{h_im_j
                    -\frac{3}{4}\sum_Ik^I{}_il_{Ij}}{z_{ji}} +c\right)d\phi.
\end{eqnarray}
The asymptotic flatness requires that $f^{-1}$ approaches to $1$ and $\omega$ vanishes at infinity in the given coordinate system. 
Therefore, we impose the following conditions on the parameters $(l_{I0},c,m_0)$
\begin{eqnarray}
l_{I0}&=&\frac{1}{3},\label{eq:l0}\\
c&=&-\sum_{i,j(i\not =j)}\frac{h_im_j-\frac{3}{4}\sum_Ik^I{}_il_{Ij}}{z_{ji}},\label{eq:c}\\
m_0&=&\frac{3}{4}\sum_{i,I}k^I{}_il_{I0}.\label{eq:m0}
\end{eqnarray}
In terms of the five-dimensional radial coordinate $\rho:=2\sqrt{r}$, for $\rho \to \infty\ (r\to \infty)$ the metric is asymptotically approximated as
\begin{eqnarray}
ds^2&\simeq& -dt^2+d\rho^2+\frac{\rho^2}{4}\left[(d\psi+\cos\theta d\phi)^2+ d\theta^2+\sin^2\theta d\phi^2 \right],
\end{eqnarray}
which coincides with the metric of Minkowski spacetime written in terms of the Euler angles. 
The regularity of the metric at infinity demands the range of the coordinates $0\le \theta\le \pi$, $0\le \phi<2\pi$, $0\le \psi<4\pi$ and 
the identification of $\phi\sim \phi+2\pi$ and $\psi\sim\psi+4\pi$.




\subsection{Horizon ${\bm r}={\bm r}_1$}
To see that the point ${\bm r}={\bm r}_1$ denotes a Killing horizon with the topology of the lens space of $L(n,1)=S^3/{\mathbb Z}_n$, 
let us introduce new spherical polar coordinates such that ${\bm r}={\bm r}_1$ becomes  an origin in ${\mathbb E}^3$ of the Gibbons-Hawking space. 
Near the origin $r:=|{\bm r}-{\bm r}_1|=0$, the functions $f^{-1}$ and $\omega_\psi$ are approximated as 
\begin{eqnarray}
f^{-1}&\simeq& \frac{P}{4h_1r}+c_1',\\
\omega_\psi &\simeq&-\frac{k^1{}_{1}k^2{}_{1}k^3{}_{1}+6h_1\sum_Ik^I{}_1l_{I1}-8h_1^2m_1}{8h_1^2r}+c_2',
\end{eqnarray}
where we have defined the constants $c_1'$ and $c_2'$, respectively,  by
\begin{eqnarray}
c_1'&:=&\frac{1}{3P^2}\left[k^1{}_1k^2{}_1k^3{}_1\sum_Ik^I{}_1+72h_1^2\sum_{I,J,K}|\epsilon_{IJK}|l_{I1}l_{J1}+6h_1\sum_{I,J,K}|\epsilon_{IJK}|k^I{}_1k^J{}_1(l_{I1}+l_{J1})\right]\notag\\
&+&\frac{1}{12P^2h_1^2}\sum_{i\not=1}\frac{1}{|z_{i1}|}
\Biggl[-3\biggl(k^1{}_1k^2{}_1k^3{}_1)^2+8h_1k^1{}_1k^2{}_1k^3{}_1\sum_Ik^I{}_1l_{I1}\notag\\
&&+24h_1^2\sum_{I,J,K}|\epsilon_{IJK}|k^I{}_1k^J{}_1l_{I1}l_{J1}\biggr)h_i+h_1\sum_{I}\left(\sum_{J,K}|\epsilon_{IJK}|k^J{}_1k^K{}_1+24h_1l_{I1}\right)\notag\\
&\times&\left(k^1{}_1k^2{}_1k^3{}_1+3h_1\sum_{J,K}|\epsilon_{IJK}|(k^J{}_1l_{J1}+k^K{}_1l_{K1})\right)k^I{}_i\notag \\
&+&6h_1^2\sum_{I,J,K}|\epsilon_{IJK}|(k^I{}_1k^K{}_1+12h_1l_{J1})(k^I{}_1k^J{}_1+12h_1l_{K1})l_{Ii}
\Biggr],\\
c_2'&:=&m_0-\frac{1}{4h_1}\sum_Ik^I{}_1+\sum_{i\not =1}\frac{1}{8h_1^3|z_{i1}|}\Biggl[\left(2k^1{}_1k^2{}_1k^3{}_1+6h_1\sum_Ik^{I1}l_{I1}\right)h_i\notag\\
&&-\frac{h_1}{2}\sum_I\left(\sum_{J,K}|\epsilon_{IJK}|k^J{}_1k^K{}_1+12h_1l_{I1}\right)k^I{}_i-6h_1^2\sum_I k^I{}_1l_{Ii}+8h_1^3m_i\Biggr],
\end{eqnarray}
with
\begin{eqnarray}
P:=\left\{ (12h_1l_{11}+k^2{}_1k^3{}_1)(12h_1l_{21}+k^3{}_1k^1{}_1)(12h_1l_{31}+k^1{}_1k^2{}_1) \right\}^{\frac{1}{3}}.
\end{eqnarray}
Moreover, to eliminate the apparent divergence of the metric at $r=0$, let us use new coordinates $(v,\psi')$ defined by
\begin{eqnarray}
dv=dt-\left(\frac{A_0}{r^2}+\frac{A_1}{r}\right)dr,\qquad 
d\psi'=d\psi+\sum_{i\not=1}\frac{z_{i1}}{|z_{i1}|}d\phi-\frac{B_0}{r}dr.
\end{eqnarray}
where
\begin{eqnarray}
4A_0&=&\biggl[ -9(k^1{}_1l_{11}-k^2{}_1l_{21})^2+18l_{31}(k^1{}_1k^3{}_1l_{11}+k^2{}_1k^3{}_1l_{21}+24h_1l_{11}l_{21}) -9(k^3{}_1l_{31})^2\notag\\
&+&4\left(k^1{}_1k^2{}_1k^3{}_1+6h_1\sum_Ik^I{}_1l_{I1}\right)m_1-16h_1^2m_1^2\biggr]^{\frac{1}{2}},\\
-8A_0B_0&=&k^1{}_1k^2{}_1k^3{}_1+6h_1\sum_Ik^I{}_1l_{I1}-8h_1^2m_1,\\
16A_0A_1&=&
\frac{3}{2}\sum_{I,J,K}|\epsilon_{IJK}|(-k^I{}_1+k^J{}_1+k^K{}_1)k^I{}_1l_{I1}
+36h_1\sum_{I,J,K}|\epsilon_{IJK}|l_{J1}l_{K1}\notag\\
&+&2\biggl( k^1{}_1k^2{}_1k^3{}_1+6h_1\sum_{I}k^I{}_1l_{I1} \biggr)m_0
+4h_1\left(\sum_Ik^I{}_1-4h_1m_0\right)m_1\notag\\
&+&\sum_{j\ge 2}\frac{1}{|z_{j1}|}\biggl[
4\biggl( 54l_{11}l_{21}l_{31}+3\sum_{I}k^I{}_1l_{I1}m_1-4h_1m_1^2  \biggr)h_j \notag\\
&+&\sum_I\Biggl[\frac{9}{2}\sum_{J,K}\left(-k^I{}_1l_{I1}+k^J{}_1l_{J1}+k^K{}_1l_{K1}\right)l_{I1}
+\left(\sum_{J,K}|\epsilon_{IJK}|k^K{}_1k^J{}_1+12h_1l_{I1}\right)m_1\Biggr]k^I{}_j \notag \\
&+&\sum_I\Biggl[\frac{9}{2}\sum_{J,K}\left(-k^I{}_1l_{I1}+k^J{}_1l_{J1}+k^K{}_1l_{K1}\right)k^I{}_1+12h_1\sum_{J,K}|\epsilon_{IJK}|k^K{}_1k^J{}_1+12h_1k^I{}_1m_1\Biggr]l_{Ij}  \notag\\
&+&2\biggl( k^1{}_1k^2{}_1k^3{}_1+6h_1\sum_{I}k^I{}_1l_{I1}-8h_1^2m_1 \biggr)m_j   \biggr].
\end{eqnarray}
It is shown from $g_{vv}={\cal O}(r)$ that $r=0$ is a null Killing horizon. 
We can obtain the near-horizon geometry by putting $(v,r)=(v/\epsilon,\epsilon r)$ and taking the limit $\epsilon\to 0$ as 
\begin{eqnarray}
ds^2_{NH}&=&\frac{R_2^2}{4}\left[d\psi'+n\cos\theta d\phi-\frac{16A_0B_0}{R_1^2R_2^2}rdv\right]^2\nonumber\\
               &+&R_1^2(d\theta^2+\sin^2\theta d\phi^2)-\frac{4r^2}{R_1^2 R_2^2}dv^2-\frac{4}{R_2}dvdr,
\end{eqnarray}
where
\begin{eqnarray}
R_1^2&:=&\frac{P}{4},\\
R_2^2&:=&\frac{P^3-\left[k^1{}_1k^2{}_1k^3{}_1+6h_1\sum_Ik^I{}_1l_{I1}-8h_1^2m_1 \right]^2}{16h_1^2R_1^4}.
\end{eqnarray}
As expected, this is locally isometric to the near-horizon geometry of the BMPV black hole. 
The metric of the cross section of the event horizon with $v,r=$const. surfaces gives 
\begin{eqnarray}
ds^2_H=\frac{R_2^2}{4}\left[d\psi'+n\cos\theta d\phi\right]^2+R_1^2(d\theta^2+\sin^2\theta d\phi^2),
\end{eqnarray}
which is the metric of the squashed lens space $L(n,1)=S^3/{\mathbb Z}_n$. 
To remove CTCs near the horizon, we must impose
\begin{eqnarray}
R_1^2>0,\quad R_2^2>0. \label{eq:R1R2ineq}
\end{eqnarray}
Note that it is sufficient to impose the second inequality only, namely, 
\begin{eqnarray}
P>\left[k^1{}_1k^2{}_1k^3{}_1+6h_1\sum_Ik^I{}_1l_{I1}-8h_1^2m_1 \right]^\frac{2}{3}. \label{eq:R2ineq}
\end{eqnarray}




\subsection{${\bm r}={\bm r}_i$ ($n=2,...,n$)}
Next, we introduce new spherical polar coordinates such that ${\bm r}={\bm r}_i\ (n=2,...,n)$ becomes an origin in ${\mathbb E}_3$ of the Gibbons-Hawking space. 
Near the points $r=|{\bm r}-{\bm r}_i|$ ($n=2,...,n$), the functions $f^{-1}$ and $\omega_\psi$ behaves, respectively,  as
 \begin{eqnarray}
f^{-1}&\simeq& \frac{\{3(12l_{1i}-k^2{}_ik^3{}_i)(12l_{2i}-k^3{}_ik^1{}_i)(12l_{3i}-k^1{}_ik^2{}_i)\}^{\frac{1}{3}}}{12r}+c_{1(i)},\\
\omega_\psi &\simeq&\frac{-k^1{}_ik^2{}_ik^3{}_i+6(k^1{}_il_{1i}+k^2{}_il_{2i}+k^3{}_il_{3i})+8m_i}{8r}+c_{2(i)},
\end{eqnarray}
where the constants $c_{1(i)}$ and $c_{2(i)}$ $(i=2,\ldots,n)$ are given by
\begin{eqnarray}
c_{1(i)}&:=&-\frac{1}{4}\prod_{I,J(I<J)}\left( -4+\sum_{j(\not= i)}\frac{k^I{}_ik^J{}_ih_{j}+k^J{}_ik^I{}_j+k^I{}_ik^J{}_j-12\sum_{K}\epsilon_{IJK}l_{Kj}}{|z_{ji}|}  \right)^{\frac{1}{3}},\\
c_{2(i)}&:=&m_0+\frac{1}{4}\sum_Ik^I{}_i\notag\\
&&+\sum_{j\not=i}\frac{-k^1{}_ik^2{}_ik^3{}_ih_j-\frac{1}{2}\sum_{I,J,K}|\epsilon_{IJK}|k^I{}_ik^J{}_ik^K{}_j+12\sum_Ik^I{}_il_{Ij}+16m_j}{16|z_{ji}|}.
\end{eqnarray}
The $1$-forms $\chi$ and $\hat\omega$ are approximated, respectively, as
\begin{eqnarray}
\chi&\simeq& (-\cos\theta +\chi_{(0)})d\phi,\\
\hat\omega &\simeq& (\hat\omega_{(1)}\cos\theta+\hat\omega_{(0)})d\phi, \label{eq:omega}
\end{eqnarray}
where 
\begin{eqnarray}
\chi_{(0)}&=&-\sum_{j(\not=i)}\frac{h_jz_{ji}}{|z_{ji}|},\\
\hat\omega_{(0)} &:=&\sum_{k,j(k\not=j,k,j\not=i)}\left( h_km_j-\frac{3}{4}\sum_Ik^I{}_kl_{Ij} \right)\frac{z_{ji}z_{ki}}{z_{jk}|z_{ji}||z_{ki}|}\notag\\
&&-\sum_{j(\not=i)}\left(m_0h_j-\frac{3}{4}\sum_Ik^I{}_jl_{I0}\right)\frac{-z_{ji}}{|z_{ji}|}+c,\\
\hat\omega_{(1)} &:=&\sum_{j(j\not=i)}\left[h_im_j-h_jm_i-\frac{3}{4}\sum_I(k^I{}_il_{Ij}-k^I{}_jl_{Ii})\right]\frac{-1}{|z_{ji}|}-\left(m_0h_i-\frac{3}{4}\sum_Ik^I{}_il_{I0}\right).
\end{eqnarray}
To eliminate the divergence at the points ${\bm r}={\bm r}_i$ ($n=2,...,n$) of the function $f^{-1}$ and $\omega_\psi$, we need impose the following conditions on the parameters $(l_{Ii},m_{i})\ (i=2,\ldots, n,I=1,2,3)$
\begin{eqnarray}
l_{1i}=\frac{k^2{}_ik^3{}_i}{12},\quad l_{2i}=\frac{k^3{}_ik^1{}_i}{12},\quad l_{3i}=\frac{k^1{}_ik^2{}_i}{12}, \label{eq:condi_l}
\end{eqnarray}
\begin{eqnarray}
m_i=-\frac{k^1{}_ik^2{}_ik^3{}_i}{16}. \label{eq:condi_m}
\end{eqnarray}
This immediately leads to the important constraint
\begin{eqnarray}
\hat\omega_{(1)}=c_{2(i)}, \ (i=2,\ldots,n).\label{eq:omega1}
\end{eqnarray}
Using these conditions and new coordinates defined by
\begin{eqnarray}
\rho=\sqrt{-c_{1(i)}r},\quad \psi'=\psi+\chi_{(0)}\phi,\quad \phi'=\phi, \label{eq:coordinate}
\end{eqnarray}
we find that near the points ${\bm r}={\bm r}_i\ (i=2,\ldots,n)$ the metric is locally isometric to the Minkowski metric 
\begin{eqnarray}
ds^2\simeq -c_{1(i)}^{-2}d(t+c_{2(i)}\psi'+\hat\omega_{(0)}\phi')^2+d\rho^2+\frac{\rho^2}{4}\{(d\psi'-\cos\theta d\phi')^2+d\theta^2+\sin^2\theta d\phi'^2\}.
\end{eqnarray}
However,  the existence of $c_{2(i)}$ and $\hat\omega_{(0)}$ in the metric yields CTCs around the origin $\rho=0$ (${\bm r}={\bm r}_i\ (i=2,\ldots,n)$). 
Therefore, we  must require 
\begin{eqnarray}
c_{2(i)}=0,\quad \hat\omega_{(0)}=0,\label{eq:c20}
\end{eqnarray}
where it should be noted that the former automatically guarantees the latter. 
Moreover, from Eq.(\ref{eq:coordinate}), we must require $c_{1(i)}<0\ (i=2,\ldots,n)$, which can be written as
\begin{eqnarray}
\prod_{I,J(I<J)}\left( -4+\sum_{j(\not= i)}\frac{k^I{}_ik^J{}_ih_{j}+k^J{}_ik^I{}_j+k^I{}_ik^J{}_j-12\sum_{K}\epsilon_{IJK}l_{Kj}}{|z_{ji}|}  \right)>0. \label{eq:c1ineq}
\end{eqnarray}




\subsection{Axis}
On ${\mathbb E}^3$  in the Gibbons-Hawking space (Eq.~(\ref{eq:ds3})), let us introduce the Cartesian coordinates $(x,y,z)$, which are defined by $(x,y,z)=(r\sin\theta \cos\phi,r\sin\theta\sin\phi,r\cos\theta)$.

First, we show that there exist no Dirac-Misner string singularities on the $z$-axis of ${\mathbb E}^3$ (i.e., $x=y=0$) in the Gibbons-Hawking space. To do so, it is sufficient to show $\hat\omega=0$ on the $z$-axis.
The $z$-axis is split into  the $(n+1)$ intervals as $I_-=\{(x,y,z)|x=y=0,  z<z_1\}$, $I_i=\{(x,y,z)|x=y=0,z_i<z<z_{i+1}\}\ (i=1,...,n-1)$ and $I_+=\{(x,y,z)|x=y=0,z>z_n\}$. 
 On the $z$-axis, the 1-form $\hat \omega$ can be written in the form
\begin{eqnarray}
\hat\omega
&= &\sum_{k,j(k\not=j)}\left(h_km_j-\frac{3}{4}\sum_Ik^I{}_kl_{Ij}\right)\frac{(z-z_k)(z-z_j)}{z_{jk}|z-z_k||z-z_j|}d\phi\notag\\
&-&\sum_{j}\left(m_0h_j-\frac{3}{4}\sum_Ik^I{}_jl_{I0}\right)\frac{z-z_j}{|z-z_j|}d\phi -\sum_{k,j(k\not=j)}\left(h_km_j-\frac{3}{4}\sum_Ik^I{}_kl_{Ij}\right)\frac{d\phi}{z_{jk}},
\end{eqnarray}
where we have used Eq.~(\ref{eq:c}). 
On $I_\pm$, $\hat\omega$ vanishes since 
\begin{eqnarray}
\hat\omega=\mp\sum_{j}\left(m_0h_j-\frac{3}{4}\sum_Ik^I{}_jl_{I0} \right)d\phi=0,
\end{eqnarray}
where we have used  Eq.~(\ref{eq:m0}).
For $z\in I_i\ (i=1,2,...,n-1)$, 
we find
\begin{eqnarray}
\hat \omega_{\phi}|_{I_i}&=&\hat \omega_{\phi}(x=y=0,z_i<z<z_{i+1}) \\
                                       &=&\hat \omega_{\phi}(x=y=0,z=z_i+0) \\
                                        &=&\hat \omega_{\phi}(r:=|{\bm r}-{\bm r}_i|=0,\theta=0)\\
                                        &=&\hat \omega_{(1)}+\hat \omega_{(0)}\\
                                        &=&0 \label{eq:homega0}
\end{eqnarray}
where we have used the fact that $\hat\omega$ is constant on $I_i$ in the second equality and  Eq.~(\ref{eq:omega}) in the fourth equality, respectively. 
Furthermore, we have used Eq.~(\ref{eq:omega1}) and Eq.~(\ref{eq:c20}) in the last equality. 
Thus, it has been shown that $\hat\omega=0$ holds at each interval, which proves that no Dirac-Misner string pathologies exist throughout the spacetime. 
\medskip

Next, we show that there exist no orbifold singularities on the $z$-axis. 
On $I_\pm$, $\chi$ can be written as
\begin{eqnarray}
\chi&=&\pm d\phi,
\end{eqnarray} 
whereas on $I_i$ we have 
\begin{eqnarray}
\chi
      &=&\left(n\frac{z-z_1}{|z-z_1|}-\sum_{2\le j\le i}\frac{z-z_j}{|z-z_j|}-\sum_{i+1\le j\le n-1}\frac{z-z_j}{|z-z_j|}\right)d\phi \notag \\
      &=&\left(2n-2i+1\right)d\phi.
\end{eqnarray}
Let us use  the coordinate basis vectors $(\partial_{\phi_1},\partial_{\phi_2})$ of $2\pi$ periodicity, which are defined by $\phi_1:=(\psi+\phi)/2$ and $\phi_2:=(\psi-\phi)/2$.  
We can show that  the Killing vector $v:=\partial_\phi-\chi_\phi\partial_\psi$ vanishes on each interval. 
Therefore, the rod structure is given by
\begin{itemize}
\item on $I_+$, the Killing vector $v_+:=\partial_\phi-\partial_\psi=(0,-1)$ vanishes,
\item on each  $I_i$ ($i=1,...,n-1$), the Killing vector $v_i:=\partial_\phi-(2n-2i+1)\partial_\psi=(i-n,i-n-1)$ vanishes,
\item on  $I_-$, the Killing vector $v_-:=\partial_\phi+\partial_\psi=(1,0)$ vanishes. 
\end{itemize}
From these, we can observe that the Killing vectors $v_\pm,\ v_i$ on the intervals satisfy 
\begin{eqnarray}
{\rm det}\ (v_+^T,v_{n-1}^T)=-1,\qquad {\rm det}\ (v_{i}^T,v_{i-1}^T)=-1, 
\label{noorbifold}
\end{eqnarray}
with 
\begin{eqnarray}
{\rm det}\ (v_1^T,v_{-}^T)=n.
\label{lens_cond}
\end{eqnarray}
Eq. (\ref{noorbifold}) means that  there exist no orbifold singularities at adjacent intervals $z=z_i\ (1\le i\le n)$, and Eq. (\ref{lens_cond}) shows that  the spatial topology of the horizon is the lens space $L(n,1)=S^3/{\mathbb Z}_n$.




\section{Physical properties}
\label{sec:analysis}
 Let us count the number of physical  degree of freedom.
First, note that there exists a gauge freedom of redefining harmonic functions
\begin{eqnarray}
&&K^I \to K^I-a^I H,\notag \\
&&L_I \to L_I -\frac{1}{12}|\epsilon_{IJK}|a^J K^K-\frac{1}{24}|\epsilon_{IJK}| a^Ja^K H,\label{eq:gauge}\\
&&M\to M-\frac{3}{4} a^I L_I+\frac{1}{96}|\epsilon_{IJK}|(a^Ia^Ja^KH-3a^Ia^J K^K), \notag
\end{eqnarray}
where $a^I$ are three arbitrary constants. 
Under these transformations,  $f, H_I,X^I,\omega, \chi$ remain invariant. 
The 1-forms $\xi^I$ transforms as $\xi \to \xi +a^I\chi $, which means merely the gauge transformation since $A^I \to A^I-a^Id\psi$. 
Therefore, the transformations~(\ref{eq:gauge}) makes the solution invariant. 
This gauge invariance enables us to eliminate one term which appears in $K^I$ and to simplify the form of the solution. 
We can use this gauge invariance to put, for instance, 
  \begin{eqnarray}
k^I{}_1=0. \label{eq:k10}
\end{eqnarray}

\medskip
The regularity of the metric at each boundary has required the boundary conditions (\ref{eq:l0})-(\ref{eq:m0}), (\ref{eq:condi_l})-(\ref{eq:condi_m}), (\ref{eq:c20}).   
From the gauge freedom of $z\to z+z_0\ (z_0: {\rm constant})$ and (\ref{eq:k10}), these conditions reduce the number of the independent parameters that appear in the solution from $(8n+5)$ to $(3n+1)$, 
Furthermore, the absence of CTCs  demands that these should satisfy the inequalities (\ref{eq:R2ineq}) and  (\ref{eq:c1ineq}).

We compute the conserved charges of the black lens. 
The Arnowitt-Deser-Misner (ADM) mass, the electric charges and two ADM angular momenta are written as 
\begin{eqnarray}
M&=&\sum_IQ_I=\sum_I\frac{3\pi }{G}\left(   \sum_{i} l_{Ii}+\frac{1}{24} \sum_{\substack{J,K\\ i,j}}    |\epsilon_{IJK}| k^J{}_ik^K{}_j                          \right),\\
J_\phi&=&\frac{\pi}{2G}  \sum_{\substack{I \\i,j}}   \left( h_i k^I{}_j - h_j k^I{}_i               \right)z_i,\\
J_\psi&=&\frac{\pi}{G}\left[   \frac{1}{24} \sum_{\substack{I,J,K\\ i,j,k}} |\epsilon_{IJK}|  k^I{}_ik^J{}_j k^K{}_k  +\frac{3}{2} \sum_{\substack{I\\ i,j  }}k^I{}_i l_{Ij} -2\sum_i m_i \right].
\end{eqnarray}
As expected, the mass saturates the Bogomol’nyi-Prasad-Sommerfield (BPS) bound $M\ge \sum_IQ_I$.
The magnetic fluxes through $I_i\ (i=1,\ldots,n-1)$ are defined as $q[I_i]:=\frac{1}{4\pi}\int_{I_i} F$, which  are computed as
\begin{eqnarray}
q^I[I_1]&=&\frac{k^1{}_1k^2{}_1k^3{}_1+6nk^I{}_1 l_{I1}-8n^2m_1}{2n(24nl_{I1}+k^I{}_1 k^K{}_1|\epsilon_{IJK}|)}+\frac{1}{2}\left(\frac{k^I{}_1}{n}+k^I{}_2\right),\\
q^I[I_i]&=&\frac{1}{2}(k^I{}_{i}-k^I{}_{i+1}).
\end{eqnarray}




\section{Summary}
\label{sec:discuss}
In this paper,  we have generalized the asymptotically flat supersymmetric black lens solution with the horizon topology of $L(n,1)$ in the five-dimensional minimal supergravity~\cite{Tomizawa:2016kjh} to the five-dimensional $U(1)^3$ supergravity, which also corresponds to the extension of the black lens solution with the lens space $L(2,1)$ to the black lens solution with the more genera lens spaces $L(n,1)\ (n\ge 3)$. 
We have shown that the black lens with the horizon  topology of $L(n,1)$ includes $3n+2$ parameters, which must obey $n$ inequalities. 
We have computed $(3n+1)$ physical charges, three electric charges $Q_I\ (I=1,2,3)$ (the sum is equal to the mass, and hence it follows that the Bogomol’ny bound is saturated), two angular momenta $J_\phi,J_\psi$, and $(n-1)$ magnetic fluxes $q^I{}_i\ (i=1,\ldots,n-1)$, which are subject to a constraint.

\medskip
In this work, we have impose the boundary conditions such that the point source ${\bm r}={\bm r}_1$ in the harmonic functions corresponds to the Killing horizon and the other ${\bm r}={\bm r}_i\ (i=2,\ldots,n)$ are simply regular points like an origin of Minkowski spacetime. 
It may be also interesting to consider the different boundary conditions, such that ${\bm r}={\bm r_i}\ (2\le i \le [\frac{n+1}{2}])$ is a Killing horizon and the other are regular points, since such a solution is expected to have physically different properties (see Ref.~\cite{Kunduri:2014kja} for the corresponding examples in five-dimensional  minimal supergravity). 
Furthermore, it may be straightforward to generalize the present solution to five-dimensional $U(1)^N$ supergravity. 
These issues deserve further study.




\acknowledgments
This work was supported by the Grant-in-Aid for Scientific Research (C) (Grant Number ~17K05452) from the Japan Society for the Promotion of Science.





\begin{thebibliography}{99}



 \bibitem{Hollands:2007aj} 
  S.~Hollands and S.~Yazadjiev,
  ``Uniqueness theorem for 5-dimensional black holes with two axial Killing fields,''
  Commun.\ Math.\ Phys.\  {\bf 283}, 749 (2008)
  [arXiv:0707.2775 [gr-qc]].
  
\bibitem{Hollands:2010qy} 
  S.~Hollands, J.~Holland and A.~Ishibashi,
  ``Further restrictions on the topology of stationary black holes in five dimensions,''
  Annales Henri Poincare {\bf 12}, 279 (2011)
   [arXiv:1002.0490 [gr-qc]].
   
   \bibitem{Cai:2001su} 
  M.~l.~Cai and G.~J.~Galloway,
  ``On the Topology and area of higher dimensional black holes,''
  Class.\ Quant.\ Grav.\  {\bf 18}, 2707 (2001)
  [hep-th/0102149].
 

\bibitem{Galloway:2005mf} 
  G.~J.~Galloway and R.~Schoen,
  ``A Generalization of Hawking's black hole topology theorem to higher dimensions,''
  Commun.\ Math.\ Phys.\  {\bf 266}, 571 (2006)
  [gr-qc/0509107].

\bibitem{Myers:1986un} 
  R.~C.~Myers and M.~J.~Perry,
  ``Black Holes in Higher Dimensional Space-Times,''
  Annals Phys.\  {\bf 172}, 304 (1986).
 
\bibitem{Emparan:2001wn} 
  R.~Emparan and H.~S.~Reall,
  ``A Rotating black ring solution in five-dimensions,''
  Phys.\ Rev.\ Lett.\  {\bf 88}, 101101 (2002)
  [hep-th/0110260].
 
\bibitem{Pomeransky:2006bd} 
  A.~A.~Pomeransky and R.~A.~Sen'kov,
  ``Black ring with two angular momenta,''
  hep-th/0612005.
  
  
  
  
   
\bibitem{Evslin:2008gx} 
  J.~Evslin,
  ``Geometric Engineering 5d Black Holes with Rod Diagrams,''
  JHEP {\bf 0809}, 004 (2008)  [arXiv:0806.3389 [hep-th]].
 
\bibitem{Chen:2008fa}
Y.~Chen and E.~Teo, ``A Rotating black lens solution in five dimensions'', {} Phys.\ Rev.\ D {\bf 78}, 064062 (2008).




\bibitem{Tomizawa:2019acu}
  S.~Tomizawa and T.~Mishima,
  ``A stationary and biaxisymmetric four-soliton solution in five dimensions,''
  arXiv:1902.10544 [hep-th].



\bibitem{Gauntlett:2002nw} 
  J.~P.~Gauntlett, J.~B.~Gutowski, C.~M.~Hull, S.~Pakis and H.~S.~Reall,
  ``All supersymmetric solutions of minimal supergravity in five- dimensions,''
  Class.\ Quant.\ Grav.\  {\bf 20}, 4587 (2003) 
  [hep-th/0209114].
 



  \bibitem{Reall:2002bh} 
  H.~S.~Reall,
  ``Higher dimensional black holes and supersymmetry,''
  Phys.\ Rev.\ D {\bf 68}, 024024 (2003)
  Erratum: [Phys.\ Rev.\ D {\bf 70}, 089902 (2004)]
  [hep-th/0211290].
  


  
  

  \bibitem{Breckenridge:1996is} 
  J.~C.~Breckenridge, R.~C.~Myers, A.~W.~Peet and C.~Vafa,
  ``D-branes and spinning black holes,''
  Phys.\ Lett.\ B {\bf 391}, 93 (1997)
  [hep-th/9602065].
  
  
  
  \bibitem{Elvang:2004rt} 
  H.~Elvang, R.~Emparan, D.~Mateos and H.~S.~Reall,
  ``A Supersymmetric black ring,''
  Phys.\ Rev.\ Lett.\  {\bf 93}, 211302 (2004)
  [hep-th/0407065].
  
    
  
  
  
  
  
  
  
  
  
  
  
  
  
  
  
  
  
  
  
  
  
  
  
  
  
    
\bibitem{Kunduri:2014kja} 
 H.~K.~Kunduri and J.~Lucietti,``Supersymmetric Black Holes with Lens-Space Topology'', {} Phys.\ Rev.\ Lett.\  {\bf 113}, no. 21, 211101 (2014).
 
\bibitem{Kunduri:2016xbo}
  H.~K.~Kunduri and J.~Lucietti,
  ``Black lenses in string theory,''
  Phys.\ Rev.\ D {\bf 94} (2016) no.6,  064007
  [arXiv:1605.01545 [hep-th]].
 
  \bibitem{Tomizawa:2016kjh} 
  S.~Tomizawa and M.~Nozawa,
  ``Supersymmetric black lenses in five dimensions,''
  Phys.\ Rev.\ D {\bf 94}, 044037 (2016) [arXiv:1606.06643 [hep-th]].


     

\bibitem{Tomizawa:2017suc}
  S.~Tomizawa and T.~Okuda,
  ``Asymptotically flat multiblack lenses,''
  Phys.\ Rev.\ D {\bf 95} (2017) no.6,  064021 [arXiv:1701.06402 [hep-th]].
  
  \bibitem{Gauntlett:2004qy}
  J.~P.~Gauntlett and J.~B.~Gutowski,
  ``General concentric black rings,''
  Phys.\ Rev.\ D {\bf 71} (2005) 045002
  doi:10.1103/PhysRevD.71.045002
  [hep-th/0408122].
   
\bibitem{Tomizawa:2017uxp} 
  S.~Tomizawa,
  ``Charged black lens in de Sitter space,''
  Phys.\ Rev.\ D {\bf 97}, no. 4, 044001 (2018)
  doi:10.1103/PhysRevD.97.044001
  [arXiv:1712.05132 [hep-th]].



























  
\end{thebibliography}
\end{document}